\documentclass[manuscript]{aastex6}
\usepackage[latin9]{inputenc}
\usepackage{array}
\usepackage{amsbsy}
\usepackage{graphicx}

\makeatletter

\providecommand{\tabularnewline}{\\}

\usepackage{color}
\usepackage{tabularx}
\usepackage{rotating}

\usepackage{array}
\shorttitle{Meridional Circulation in the Solar
Convection Zone}
\shortauthors{Pipin and Kosovichev}

\makeatother

\begin{document}

\title{On the Origin of the Double-Cell Meridional Circulation in the Solar
Convection Zone}

\author{V.V. Pipin$^{1,2}$ and A.G. Kosovichev$^{3,4,5}$}

\affil{$^{1}$National Astronomical Observatory of Japan, 2-21-1 Osawa,
Mitaka, Tokyo 181-8588, Japan\\
 $^{2}$Institute of Solar-Terrestrial Physics, Russian Academy of
Sciences, Irkutsk, 664033, Russia\\
$^{3}$Center for Computational Heliophysics, New Jersey Institute
of Technology, Newark, NJ 07102, USA\\
 $^{4}$Department of Physics, New Jersey Institute of Technology,
Newark, NJ 07102, USA\\
$^{5}$NASA Ames Research Center, Moffett Field, CA 94035, USA\\
}
\begin{abstract}
Recent advances in helioseismology, numerical simulations and mean-field
theory of solar differential rotation have shown that the meridional
circulation pattern may consist of two or more cells in each hemisphere
of the convection zone. According to the mean-field theory the double-cell
circulation pattern can result from the sign inversion of a nondiffusive
part of the radial angular momentum transport (the so-called $\Lambda$-effect)
in the lower part of the solar convection zone. Here, we show that
this phenomenon can result from the radial inhomogeneity of the Coriolis
number, which depends on the convective turnover time. We demonstrate
that if this effect is taken into account then the solar-like differential
rotation and the double-cell meridional circulation are both reproduced
by the mean-field model. The model is consistent with the
distribution of turbulent velocity correlations determined from observations
by tracing motions of sunspots and large-scale magnetic fields, indicating
that these tracers are rooted just below the shear layer. 
\end{abstract}

\section{Introduction}

Angular momentum transport on the Sun is tightly related to the influence
of the Coriolis force on motions inside the convection zone. The mean-field
hydrodynamic theory predicts that in addition to viscous redistribution
of the angular momentum by convective motions there is a nondissipative
contribution to turbulent stresses. This contribution is called the
$\Lambda$-effect \citep{KR93L}. According to this theory, the differential
rotation profile in stellar convection zones is established as a result
of a nonlinear balance between the turbulent stresses and meridional
circulation \citep{1999ApJ...511..945D}. The meridional circulation
itself is driven by perturbation of the Taylor-Proudman balance between
the centrifugal and baroclinic forces. The {baroclinic forces}
result from the nonuniform distribution of the mean entropy because of
the anisotropic heat transport in a rotating convective zone \citep{1999ApJ...511..945D}.

Along with the solar-like angular velocity profiles, the standard
mean-field models predict the meridional circulation structure with
one circulation cell in each hemisphere. Contrary, direct numerical
simulations, in most cases, predict the meridional circulation structure
with multiple cells (see, \citealt{2011ApJ...743...79M,2013Icar..225..156G,guer2013,2014AA...570A..43K,2015ApJ...804...67F,2015ASPC498.154H}).
The double-cell (or multiple-cell) structure of the meridional circulation
has been suggested by recent helioseismology inversions \citep{Zhao13m,2013ApJ778L38S,khol2014},
but is still under debate. For example, \citet{2015ApJ813.114R} found
that the meridional circulation can be approximated by a single-cell
structure with the return flow deeper than 0.77R$_{\odot}$. However,
their results indicate an additional weak cell in the equatorial region,
and contradict to the recent results of \citet{2017arXiv170708803B}
who confirmed a shallow return flow at 0.9R$_{\odot}$.

Recent mean-field hydrodynamic modeling by \citet{2016AdSpR58.1490P}
and \cite{2017ApJ835.9B} (hereafter BY17) showed that the double-cell
meridional circulation structure can be reproduced by tuning the nondissipative
turbulent stresses, i.e., the $\Lambda$-effect. In particular, BY17
argued that the double-cell meridional circulation structure can be
explained if the radial transport of angular momentum by the $\Lambda$-effect
changes sign at some depth of the convection zone. {This effect
was demonstrated by prescribing ad hoc radial profile of the $\Lambda$-tensor,
changing sign at the midpoint of the convection zone, at $r=0.825~R_{\odot}$.
Direct numerical simulations by \citet{kap2011} also showed that
the radial $\Lambda$-effect can inverse sign in a case of high Coriolis
number $\Omega^{*}=2\Omega\tau_{c},$ where $\Omega$ is the global
mean rotation rate, and $\tau_{c}$ is the local turnover time of
convection. These results motivated us to search for the turbulent
mechanism that can explain the sign inversion of the $\Lambda$-effect
in the solar convection zone.}

{In this paper, we show that the sign inversion of the radial
$\Lambda$-effect follows naturally from the standard mean-field hydrodynamics
theory \citep{KR93L,kit2004AR}. By employing the standard solar model
and the mixing-length theory of convective energy transport, we show
that the sign inversion of the $\Lambda$-effect is located in the
lower convection zone, at $r\simeq0.78~R_{\odot}$. The key point
is that in addition to the density stratification considered in the
previous mean-field models, the radial gradients of the functions
that describe effects of the Coriolis force have to be taken into
account. We demonstrate that when these gradients are taken into account
then both the solar-like distribution of the angular velocity profile
and the double-layer meridional circulation structure are both reproduced
by the mean-field model. }

\section{Basic equations.}

\subsection{The angular momentum balance}

We decompose the axisymmetric mean velocity into poloidal and toroidal
components: $\mathbf{\overline{U}}=\mathbf{\overline{U}}^{m}+r\sin\theta\Omega\hat{\mathbf{\boldsymbol{\phi}}}$,
where $\boldsymbol{\hat{\phi}}$ is the unit vector in the azimuthal
direction. The mean flow satisfies the stationary continuity equation,
\begin{equation}
\boldsymbol{\nabla}\cdot\overline{\rho}\mathbf{\overline{U}}=0,\label{eq:cont}
\end{equation}
Similarly to \citet{1989drsc.book.....R}, the conservation of the
angular momentum is expressed as follows: 
\begin{eqnarray}
\frac{\partial}{\partial t}\overline{\rho}r^{2}\sin^{2}\theta\Omega & = & -\boldsymbol{\nabla\cdot}\left(r\sin\theta\left(\hat{\mathbf{T}}_{\phi}+r\overline{\rho}\sin\theta\Omega\mathbf{\overline{U}^{m}}\right)\right)\label{eq:angm}
\end{eqnarray}
where the turbulent stresses tensor, $\hat{\mathbf{T}}$, is written
in terms of small-scale fluctuations of velocity: 
\begin{eqnarray}
\hat{T_{ij}} & = & \overline{\rho}\left\langle u_{i}u_{j}\right\rangle =\hat{T}_{i,j}^{\left(\Lambda\right)}+\hat{T}_{i,j}^{\left(\nu\right)}.\label{eq:stres}\\
 & = & \Lambda_{ijk}\overline{U}_{k}-N_{ijkl}\frac{\partial\overline{U}_{k}}{\partial r_{l}}+\dots,\label{eq:tstr}
\end{eqnarray}
{where, in following the mean-field hydrodynamic framework
(see \citealp{1989drsc.book.....R,1994AN....315..157K}), the turbulent
stress tensor is expressed as a sum of two major parts: the first
term, $\hat{T}_{i,j}^{\left(\Lambda\right)}=\Lambda_{ijk}\overline{U}_{k}$,
represents the nondissipative part (the $\Lambda$-effect), and the
second term, $\hat{T}_{i,j}^{\left(\Lambda\right)}=-N_{ijkl}\overline{U}_{k,l}$,
describes the eddy viscosity tensor contribution. The analytical expressions
for $\hat{\mathbf{T}}$ is given in the Appendix. }

To determine the meridional circulation, we consider the azimuthal
component of the large-scale vorticity, $\omega=\left(\boldsymbol{\nabla}\times\overline{\mathbf{U}}^{m}\right)_{\phi}$,
which is governed by the following equation: {
\begin{eqnarray}
\frac{\partial\omega}{\partial t} & = & r\sin\theta\boldsymbol{\nabla}\cdot\left(\frac{\hat{\boldsymbol{\phi}}\times\boldsymbol{\nabla\cdot}\overline{\rho}\hat{\mathbf{T}}}{r\overline{\rho}\sin\theta}-\frac{\mathbf{\overline{U}}^{m}\omega}{r\sin\theta}\right)\!\!+r\sin\theta\frac{\partial\Omega^{2}}{\partial z}+\frac{1}{\overline{\rho}^{2}}\left[\boldsymbol{\nabla}\overline{\rho}\times\boldsymbol{\nabla}\overline{p}\right]_{\phi}\label{eq:vort}
\end{eqnarray}
}where $\partial/\partial z=\cos\theta\partial/\partial r-\sin\theta/r\cdot\partial/\partial\theta$
is the gradient operator along the axis of rotation. Turbulent stresses
affect generation and dissipation of large-scale flows, and, in turn,
they are affected by global rotation and magnetic field. In this paper,
we neglect magnetic field effects. The magnitude of kinetic coefficients
in tensor $\hat{\mathbf{T}}$ depends on the convective turnover time,
$\tau_{c}$, and on the RMS of the convective velocity, $\mathrm{u}'$.
The radial profile of $\tau_{c}$ is obtained from the standard solar
interior model calculated using the MESA code \citep{mesa11,mesa13}.
The RMS velocity, $\mathrm{u}'$, is determined in the mixing-length
approximations from the gradient of the mean entropy, $\overline{s}$,
\[
\mathrm{u}'=\frac{\ell}{2}\sqrt{-\frac{g}{2c_{p}}\frac{\partial\overline{s}}{\partial r}},
\]
where $\ell=\mathrm{\alpha_{MLT}}H_{p}$ is the mixing length, $\alpha_{\mathrm{MLT}}=2.2$
is the mixing-length theory parameter, and $H_{p}$ is the pressure
scale height. For a nonrotating star, the $\mathrm{u}'(r)$ profile
corresponds to results of the MESA code. The mean-field equation for
the heat transport takes into account effects of rotation: 
\begin{equation}
\overline{\rho}\overline{T}\left(\frac{\partial\overline{s}}{\partial t}+\left(\overline{\mathbf{U}}\cdot\boldsymbol{\nabla}\right)\overline{s}\right)=-\boldsymbol{\nabla}\cdot\left(\mathbf{F}^{conv}+\mathbf{F}^{rad}\right)-\hat{T}_{ij}\frac{\partial\overline{U}_{i}}{\partial r_{j}},\label{eq:heat}
\end{equation}
where, $\overline{\rho}$ and $\overline{T}$ are the mean density
and temperature,and $\mathbf{F}^{conv}$ and $\mathbf{F}^{rad}$ are
the convective and radiative energy fluxes. The last term in Eq.(\ref{eq:heat})
describes the thermal energy loss and gain due to generation and dissipation
of large-scale flows. 
For the anisotropic convective flux, we employ the expression suggested
by \citet{1994AN....315..157K} (hereafter KPR94): 
\begin{equation}
F_{i}^{conv}=-\overline{\rho}\overline{T}\chi_{ij}\nabla_{j}\overline{s}.\label{conv}
\end{equation}
For calculation of the heat eddy-conductivity tensor, $\chi_{ij}$,
we take into account effects of global rotation: 
\begin{equation}
\chi_{ij}=\chi_{T}\left(\phi\left(\Omega^{*}\right)\delta_{ij}+\phi_{\parallel}\left(\Omega^{*}\right)\frac{\Omega_{i}\Omega_{j}}{\Omega^{2}}\right),\label{eq:ht-F}
\end{equation}
where functions $\phi$ and $\phi_{\parallel}$ were defined in KPR94.

The eddy conductivity and viscosity are determined from the mixing-length
approximation: 
\begin{eqnarray*}
\chi_{T} & =\mathrm{u}'\ell= & \frac{\ell^{2}}{2}\sqrt{-\frac{g}{2c_{p}}\frac{\partial\overline{s}}{\partial r}},\\
\nu_{T} & = & \mathrm{Pr}_{T}\chi_{T},
\end{eqnarray*}
where $\mathrm{Pr}_{T}$ is the turbulent Prandtl number. We found
that $\mathrm{Pr}_{T}={\displaystyle \frac{3}{4}}$ gives the magnitude
of the latitudinal differential rotation on the surface in agreement
with solar observations.

\subsubsection{{The $\Lambda$-effect profiles}}

\begin{table}
\caption{Characteristics of the mean-field hydrodynamics models.}
\centering{}%
\begin{tabular}{|c>{\centering}p{4cm}>{\centering}p{2cm}>{\centering}p{2cm}|}
\hline 
\multicolumn{1}{|c|}{Model} & \multicolumn{1}{>{\centering}p{4cm}|}{The $\Lambda$ effect} & \multicolumn{1}{>{\centering}p{2cm}}{Anisotropy parameter} & Reference\tabularnewline
\hline 
M1  & standard  &  & \citet{KR93L}\tabularnewline
\hline 
M2  & anisotropy of the backround turbulent velocity added  & a=2  & \citet{kit2004AR}\tabularnewline
\hline 
M3  & derivative of the Coriolis number added  & a=2  & this work\tabularnewline
\hline 
\end{tabular}
\end{table}

In analytical derivations of the $\Lambda$ effect within the mean-field
hydrodynamics framework, it is assumed that the background turbulence
can be modeled as a randomly forced quasi-isotropic spatially inhomogeneous
turbulent flow. Furthermore, the mean flow generation effect appears
in the second-order terms of the Taylor expansions in terms of the
large-scale inhomogeneity parameter, $\ell/L$, where $\ell$ is a
characteristic size of convective vortexes, and $L$ is a spatial
scale of mean-field parameters. In the $\Lambda$-effect calculations,
it is assumed that $L=H_{\rho}$, where $H_{\rho}=-{\displaystyle \left(\frac{\partial\log\bar{\rho}}{\partial r}\right)^{-1}}$
is the density scale height. This approximation is different from
the assumptions employed in derivations of the heat eddy-conductivity
tensor, ${\chi_{ij}}$ (see Eq.\ref{conv}), or the eddy viscosity
tensor $N_{ijkl}$. In this case, it is assumed that the background
turbulent flow is spatially homogeneous. {The dissipative effects
appear when gradients of the large-scale flow are taken into account,
i.e., additional terms of the order of $\ell/L$ in Eq(\ref{eq:tstr})
(e.g. see \citealp{1994AN....315..157K}). A detailed discussion of
approximations and assumptions of the mean-field hydrodynamics can
be found in \citet{1989drsc.book.....R}.}

{}
\begin{figure}
{\includegraphics[width=1\columnwidth]{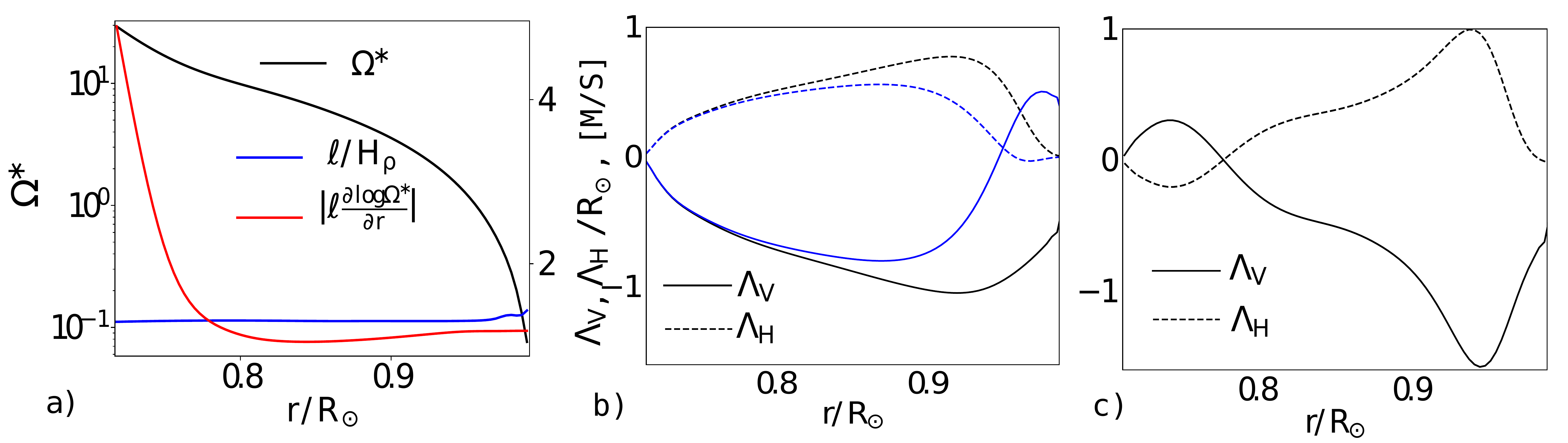} \caption{\label{fig:lamb} a) Radial profiles of parameters that affect the
magnitude and sign of the $\Lambda$-effect: the Coriolis number $\Omega^{*}=2\Omega\tau_{c}$,
where $\Omega=2.90\cdot10^{-6}$~s$^{-1}$ is the mean angular velocity
(black line), the ratio of the mixing length and the density scale
height, $\ell/H_{\rho}$ (blue line), and the ratio of the mixing
length and the scale height of the Coriolis number (red); the right vertical
axis shows the scale of the last two parameters; b) theoretical profiles
of the vertical ($\Lambda_{V}$) and horizontal ($\Lambda_{H}$) components
of the $\Lambda$-tensor in the solar convection zone at co-latitude
$\theta=45^{\circ}$ for models M1 (blue lines) and M2 (black lines)
; c) the same as b) but for model M3.}
}
\end{figure}

For accurate description of the angular momentum transfer, it is important
to take into account the effect of the radial profile of the Coriolis
number, $\Omega^{*}$, on the $\Lambda$ effect. To illustrate this,
we consider the case of fast rotation, $\Omega^{*}\gg1$. The nondiffusive
flux of angular momentum can be expressed as follows: 
\begin{eqnarray}
\Lambda_{r\phi\phi}\overline{U}_{\phi} & = & \Lambda_{V}\Omega\sin\theta,\label{eq:l1}\\
\Lambda_{V} & \approx & \nu_{T}\cos^{2}\theta\frac{J_{1}\ell}{H_{\rho}}\left(\frac{\ell}{H_{\rho}}+\ell\frac{\partial\log\Omega^{*}}{\partial r}\right),\\
\Lambda_{\theta\phi\phi}\overline{U}_{\phi} & = & \Lambda_{H}\Omega\cos\theta,\label{eq:l2}\\
\Lambda_{H} & \approx & -\nu_{T}\sin^{2}\theta\frac{J_{1}\ell}{H_{\rho}}\left(\frac{\ell}{H_{\rho}}+\ell\frac{\partial\log\Omega^{*}}{\partial r}\right),
\end{eqnarray}
where $J_{1}=-{\displaystyle \frac{\pi}{4\Omega^{*}},}$ and and $\Lambda_{V}$,
$\Lambda_{H}$ are the vertical and horizontal components of the $\Lambda$
effect. We reproduce the original results of \citet{KR93L} if the
contribution of the Coriolis number gradient, ${\displaystyle \frac{\partial\log\Omega^{*}}{\partial r}}$,
is neglected. {We would like to stress that the dependence
of the $\Lambda$ effect coefficients on the Coriolis number is included
in models M1 and M2. Those models disregard the effect of the spatial
derivatives of the $\Omega^{*}$ which seems to be important near
the bottom of the convection zone.}

However, because the convective overturn time, $\tau_{c}$, in stellar
convection zones changes with depth, the radial gradient of the Coriolis
number, $\Omega^{*}=2\Omega\tau_{c}$, can be essential for the amplitude
and direction of the $\Lambda$ effect. The $\tau_{c}$-profile can
be obtained from results of the MESA code using the standard mixing-length
approximation, $\tau_{c}=\ell/\mathrm{u}'$. Note that with the use
of the mixing-length theory for the solar interior model, this formula
is equivalent to another expression (see \citealp{1994ssebookK}):
\begin{equation}
\tau_{c}=\left(\frac{4c_{p}\overline{\rho}\ell^{2}\overline{T}}{g\delta F_{con}}\right)^{1/3},\label{eq:tauc}
\end{equation}
where $F_{con}$ is the amount of heat flux transported by convection,
and $\delta=-{\displaystyle \left(\frac{\partial\log\overline{\rho}}{\partial\log\overline{T}}\right)_{P}}$
($\delta=1$ for an ideal mono-atomic gas). The profile of the Coriolis
number calculated from the solar interior model is shown in Figure
\ref{fig:lamb}a. It is seen that parameter $|\ell{\displaystyle \frac{\partial\log\Omega^{*}}{\partial r}}|$
is greater than unity in the lower part of the convection zone, and
thus it should be taken into account.

Below, we consider results for the mean-field hydrodynamic models
of the solar differential rotation based on three models of the $\Lambda$
effect, listed in Table 1.  We would like to stress that the dependence
of the $\Lambda$ effect coefficients on the Coriolis number is included
in models M1 and M2, but these models disregard the radial derivative
of the $\Omega^{*}$. Note that models M1 and M2 differ only in the
parameter of anisotropy, which is $a=0$ in M1 and $a=2$ in M2. Model
M3 includes both the radial derivative of $\Omega^{*}$ and the convective
anisotropy.

Figures \ref{fig:lamb}(b) and (c) show the theoretical profiles of
components of the $\Lambda$-tensor in the solar convection zone.
The impact of the convective velocity anisotropy is illustrated for
the anisotropy parameter $a={\displaystyle \frac{\overline{{u_{h}^{2}}}-2\overline{{u_{r}^{2}}}}{\overline{{u_{r}^{2}}}}=}2$,
where ${u_{h}}$ and ${u_{r}}$ are the horizontal and vertical RMS
velocities. With this parameter value, the radial nondiffusive transport
of the angular momentum at the top of the convection zone is negative
(model M2). This allows us to model the subsurface shear layer. The
contribution of the ${\displaystyle \frac{\partial\log\Omega^{*}}{\partial r}}$
term, included in model M3, also changes the $\Lambda$-effect components.
These changes are of two kinds. Firstly, we see that the magnitude
of the $\Lambda$-tensor is reduced in the middle of convection zone.
Secondly, the vertical and horizontal components of the tensor both
change their sign near the bottom of the convection zone, at $r\simeq0.78~R_{\odot}$.{
}
\begin{figure}
{\includegraphics[width=1\columnwidth]{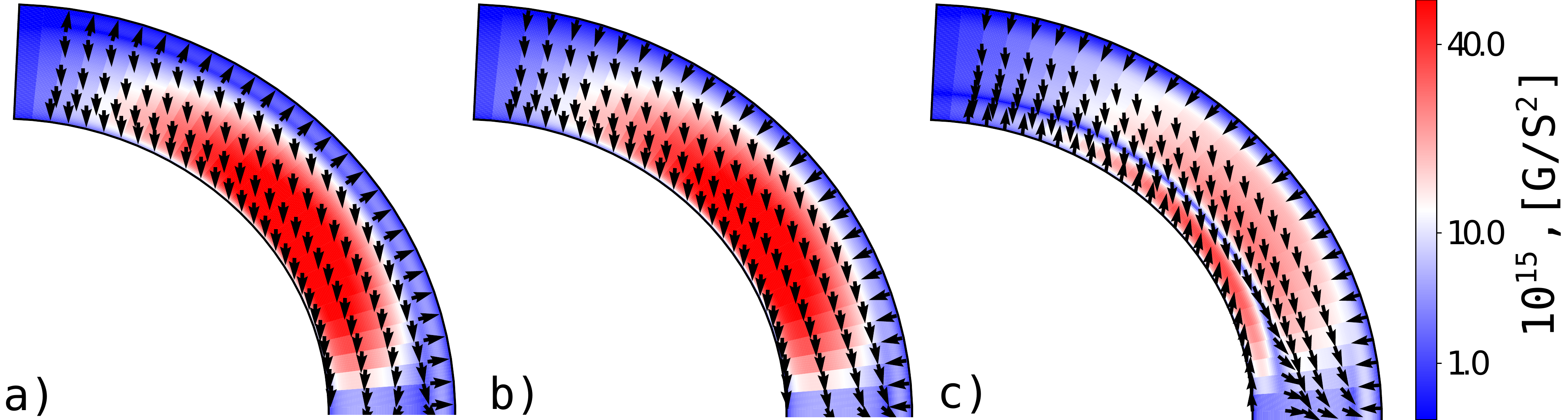}\caption{\label{fig:lam} Directions and amplitude of the angular momentum
flux due to the $\Lambda$ effect {(with the vector components
given by Eqs~(\ref{eq:l1}) and (\ref{eq:l2}))} in three models:
a) M1; b) M2; c) M3, listed in Table~1.}
}
\end{figure}

Figure \ref{fig:lam} shows the direction of the angular momentum
flux by the $\Lambda$ effect for the three models. In all of these cases,
the direction of the nondiffusive angular momentum transport is predominantly
along the axis of rotation from higher to lower latitudes in the main
part of the convection zone. Model M1 shows the outward transport
by the $\Lambda$ effect in the upper part of the convection zone.
Figure \ref{fig:lam}b (model M2) illustrates that including effects
of the convective velocity anisotropy results in the inward angular
momentum transport at the top of the convection zone. This effect
was suggested by \citet{kit2004AR}. Taking into account contributions
of the Coriolis number gradients in the $\Lambda$ effect, we get the inversion
of the angular momentum transport of direction near the base of the
convection zone (Fig. \ref{fig:lamb}c). In the upper part of the
convection zone the pattern shown in Figure \ref{fig:lam}c is in
qualitative agreement with \citet{2017ApJ835.9B} (hereafter, BY17).
Near the bottom of the convection zone both, the radial and the horizontal
components of the $\Lambda$-effect, change sign. {This is
  different from the model of BY17, who employed  an approximate fit
  of the $\Lambda$-effect profile to satisfy the gyroscopic pumping
  equation.}
{For convenience, we briefly review this concept. For details and applications,
please consult the paper by \citet{2011ApJ...743...79M}. Let us introduce
the mean specific angular momentum, $\mathcal{L}=r^{2}\sin^{2}\theta\Omega$.
Helioseismology tells that, in the solar convection zone,  $\mathcal{L}$
is constant on cylinders increasing outward from the rotational axis.
For the mean stationary stage, the equation of the angular momentum
balance can be approximated as follows: 
\begin{eqnarray}
\overline{\rho}\mathbf{\overline{U}_{\zeta}^{m}\nabla_{\zeta}\mathcal{L}} & \approx & -\boldsymbol{\nabla\cdot}\left(r\sin\theta\hat{\mathbf{T}}_{\phi}\right)\label{gyr}
\end{eqnarray}
where $\zeta=r\sin\theta$ and we take into account cylinder-like
distribution of $\mathcal{L}$.} {Note, that $\nabla_{\zeta}\mathcal{L}>0$.
The double-cell meridional circulation like that found by \citet{2014ApJ789L7Z}
gives the negative $\mathbf{\overline{U}}_{\zeta}^{m}$ at the bottom,
positive in the middle, and negative at the top of the convection
zone. Then, the LHS of Eq(\ref{gyr}) gets the same signs. Therefore,
Eq(\ref{gyr}) can be satisfied for a radially converging vector field
$\hat{\mathbf{T}}_{\phi}$. }The pattern demonstrated in Figure \ref{fig:lam}c
satisfies this condition as well as the $\Lambda$-effect model introduced
by BY17 (see Fig. 1 in their paper). As it is seen in Figure \ref{fig:lam}
that for the case of the moderate and fast rotation regimes, $\Omega^{*}>1$,
the mean-field theory predicts a nearly cylinder-like distribution
of the angular momentum fluxes produced by the $\Lambda$-effect in
the rotating stratified convective media. We note that effects of
the convective velocity anisotropy are strongly quenched with the
increase of the Coriolis number \citep{1994AN....315..157K,kit2004AR}.

\section{Results}

{}
\begin{figure}
{\includegraphics[width=1\columnwidth]{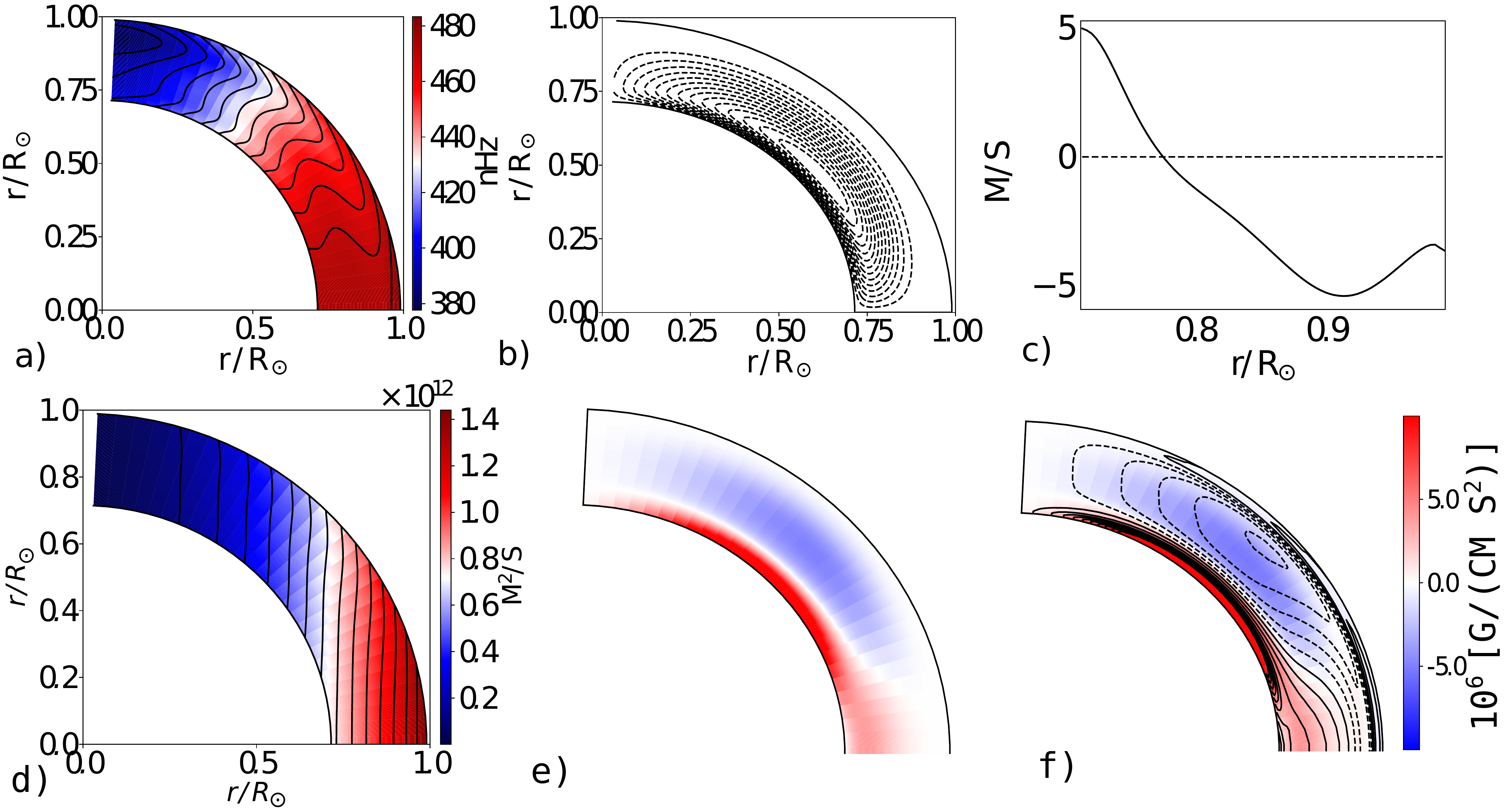} \caption{\label{fig:sun-flow00} Properties of large-scale flows in model M1
(no inversion of the $\Lambda$- effect sign and $a=0$): a) the distribution
of the rotation rate, $\Omega\left(r,\theta\right)/2\pi$, in the
solar convection zone; the isolines are plotted in the range of 385-480
nHz nHz with interval of $\approx7$nHz b) streamlines of the meridional
circulation; c) the radial profile of the meridional velocity component
at $\theta=45^{\circ}$; d) distribution of the specific angular momentum,
{$=r^{2}\sin^{2}\theta\Omega$}; e){ the angular momentum
advected by the meridional flow, i.e., $\overline{\rho}\left(\mathbf{\overline{U}}^{m}\cdot\boldsymbol{\nabla}\right)\mathcal{L}$;
f) the total torque of the turbulent stresses, $-\nabla\cdot(r\sin\theta\hat{\mathbf{T}}_{\phi})$
(background color) and the torque component from the $\Lambda$-effect
(contours), i.e., $-\nabla\cdot(r\sin\theta\hat{\mathbf{T}}_{\phi}^{\left(\Lambda\right)})$,
(see, Eqs(\ref{eq:stres},\ref{eq:tstr}) ); both are displayed in
the same interval as shown by the color bar.}}
}
\end{figure}

{Numerical solutions of Eq.~(1)-(5) for the three models of
  Table~1 are shown in Figures~3-5.} Figures \ref{fig:sun-flow00}(a)-(c)
show profiles of the angular velocity, streamlines of the meridional
circulation and the radial profile of the meridional flow velocity
at $\theta=45^{\circ}$ for model M1. The amplitude of the poleward
meridional flow velocity is of $\sim5$~m$\,$s$^{-1}$ at the surface,
and the same velocity is found near the bottom. These results are
similar to the models of \citet{1993AA...279L...1K} and \citet{1994AN....315..157K}.
Some differences with their models are likely due to the different
models of the solar interior and the $\tau_{c}$ profile. Model M1
shows a positive gradient of the angular velocity near the surface,
as found in the above cited papers, as well.{ However, this
is not consistent with results of helioseismology inversions }\citep{schouetal,Howe2011JPh}.
{Figures \ref{fig:sun-flow00}d shows distribution of the specific
angular momentum}. It has a cylinder-like profile (cf. \citealp{2011ApJ...743...79M}).
Contrary to the observations it shows the poleward deviations of the
isolines in the near-surface layer. {Figures \ref{fig:sun-flow00}
e and f , show the angular momentum advected by the meridional flow,
i.e., $\overline{\rho}\left(\mathbf{\overline{U}}^{m}\cdot\boldsymbol{\nabla}\right)\mathcal{L}$,
and the rotational forces of the $\Lambda$-effect, i.e., $-\nabla\cdot(r\sin\theta\hat{\mathbf{T}}_{\phi}^{\left(\Lambda\right)})$,
and the same for the total turbulent stress, $-\nabla\cdot(r\sin\theta\hat{\mathbf{T}}_{\phi}^{\left(\Lambda\right)})$,
(see, Eqs(\ref{eq:stres},\ref{eq:tstr}). We see that the effect
of the meridional flow is in balance with the turbulent stresses.
Contrary to results of BY17 signs of the torque from the $\Lambda$-effect
are not fully balanced with the meridional circulation. This means
that the effect of the diffusive part of the angular momentum is also
important. Our models take into account the anisotropic eddy viscosity.
Its importance for the mean-field models of the solar differential
rotations was addressed previously by \citet{1993AA...279L...1K}
and }\citet{1994AN....315..157K}{. }The mean direction of
the turbulent angular momentum flux corresponds to the direction of
the nondiffusive flux driven by the $\Lambda$- effect (see, Fig.\ref{fig:lam}a).
Note that the amplitude of the near-surface meridional circulation
is suppressed. Moreover, the model of \citet{1994AN....315..157K}
has a weak clockwise circulation cell near the surface. {Their
model, as well as model M1, has positive radial shear at the surface.
In both cases, it results from the $\Lambda$- effect profile which
corresponds to the outward angular momentum flux. }

{}
\begin{figure}
{\includegraphics[width=1\columnwidth]{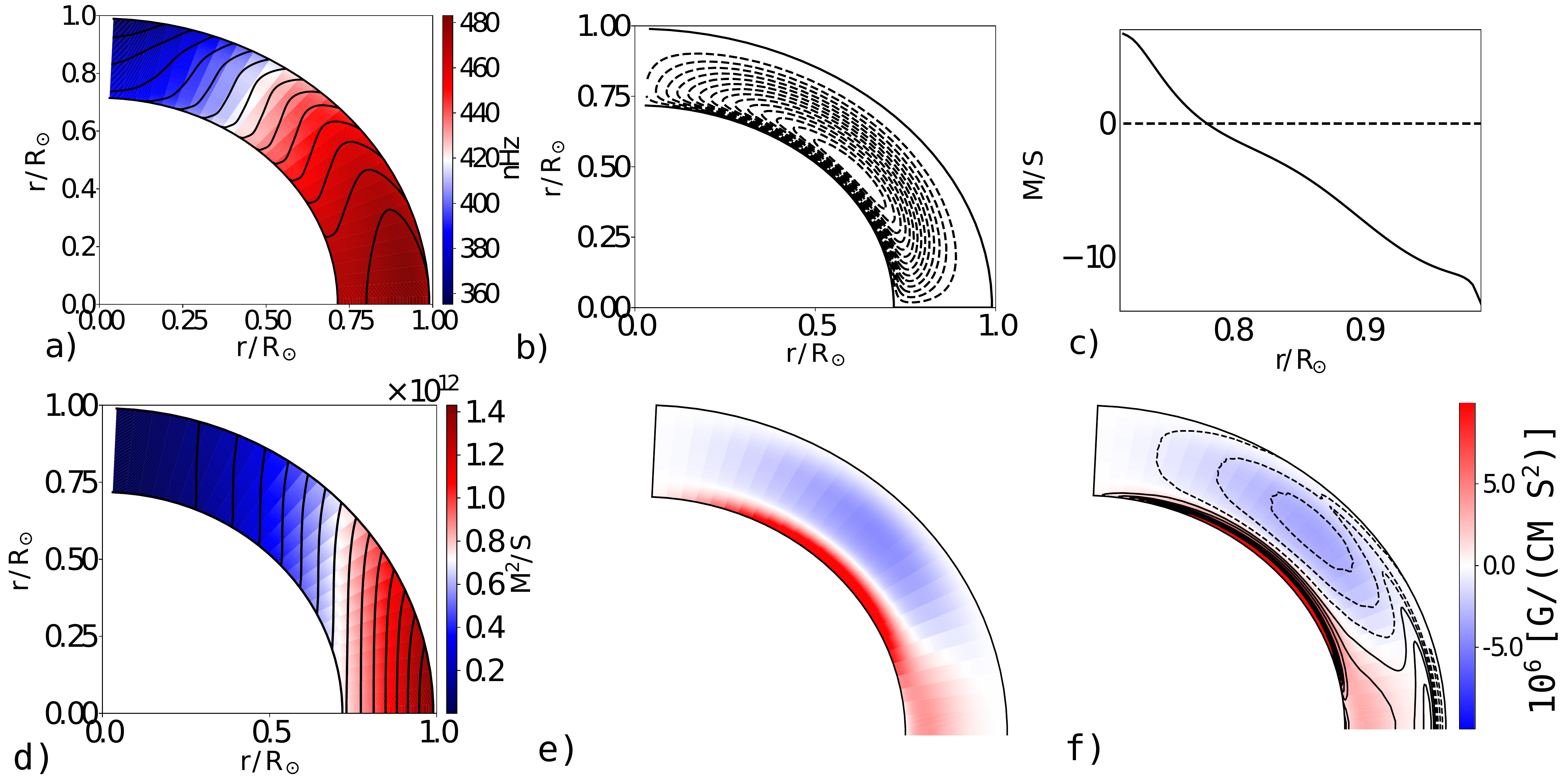} \caption{\label{fig:sun-flow0} The same as Figure \ref{fig:sun-flow00} for
model M2.}
}
\end{figure}

{Taking into account the convective velocity anisotropy in
the $\Lambda$-effect formulation, \citet{kit2004AR} and \citet{2005AN326.379K}
showed that the mean-field model can approximately reproduce the subsurface
rotational shear. Figure \ref{fig:sun-flow0} shows results for model
M2.} The results are similar to those in \citet{kit2004AR}.{
}The amplitude of the surface poleward meridional flow velocity is
about 10 m/s. {Similarly to the results of \citet{kit2004AR},
the stagnation point of the meridional circulation streamline is located
near the bottom of the convection zone. Similarly to the model M1,
the model M2 shows a balance between distributions of the angular
momentum transport by the meridional circulation and by the turbulent
stresses. This is illustrated by Figures \ref{fig:sun-flow0}e and
f. }

{}
\begin{figure}
{\includegraphics[width=1\columnwidth]{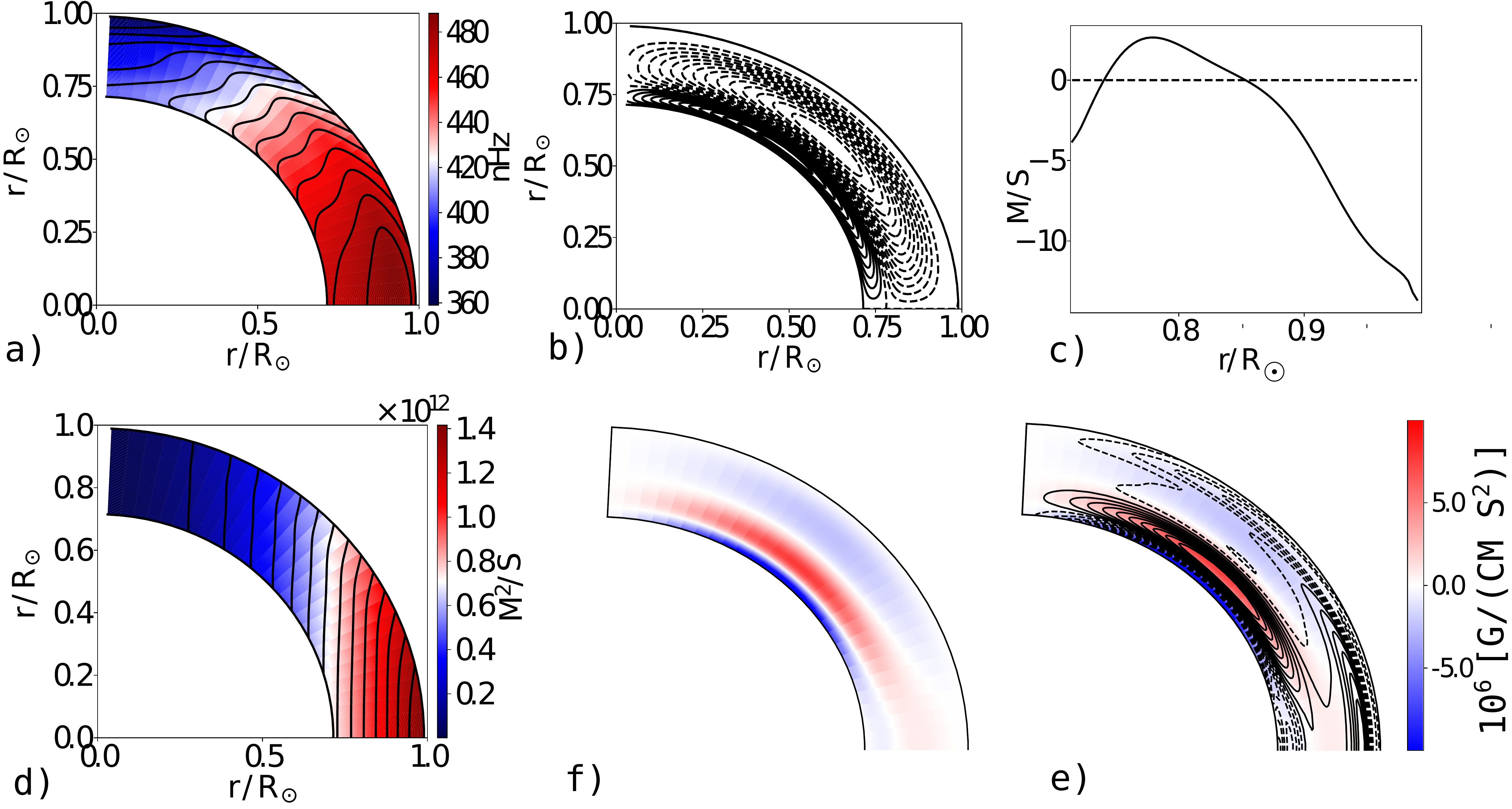} \caption{\label{fig:sun-flow}The same as Figure \ref{fig:sun-flow00} for
model M3.}
}
\end{figure}

The results for our complete model M3 (with the sign inversion of
the $\Lambda$- effect and convective anisotropy) are presented in
Fig.~\ref{fig:sun-flow}. The model shows the double-cell circulation
pattern with the upper stagnation point located at $r=0.85R_{\odot}$.
Observational results of \citet{hath12} and \citet{Zhao13m} show
it at about $0.92R_{\odot}$, and \citet{2017arXiv170708803B} found
at $0.9R_{\odot}$. The amplitude of the surface poleward flow is
about 10~m$\,$s$^{-1}$. It is likely that our model underestimates
the photospheric magnitude of the meridional flow because the top
of the integration domain is located below the photospheric level.
The angular velocity profile shows a strong subsurface shear, as well
as an increased radial gradient at the bottom.

Compared to models M1 and M2, model M3 shows a weak poleward turbulent
angular momentum flux near the bottom of the convection zone (see
Figure \ref{fig:lam}c). It results from the $\Lambda$-effect profile.{
The distribution of the angular momentum density advected by the meridional
flow, $\overline{\rho}\left(\mathbf{\overline{U}}^{m}\cdot\boldsymbol{\nabla}\right)\mathcal{L}$,
(Fig. \ref{fig:sun-flow}e) is very similar to results of BY17. Excluding
some difference in the near equatorial regions, both the model of
BY17 and our model M3 have similar distributions of the torque produced
by the $\Lambda$-effect, in particular, the negative torque at the
bottom and at the top of the convection zone. Model M3 shows the positive
torque near the equator with maximum below the subsurface shear. This
is different from BY17, and is likely due to the more complicated
structure of the $\Lambda$-effect in our model. At the bottom boundary,
the negative $\Lambda$-effect results in the positive radial shear
of the angular velocity. In high-latitude regions, the existence of
such feature disagrees with the helioseismology inversions of \citet{Howe2011JPh}.
It is likely that the issue cannot be consistently resolved without
considering dynamics of the solar tachocline. This problem is outside
the scope of this paper.}

{}
\begin{figure}
{\includegraphics[width=1\columnwidth]{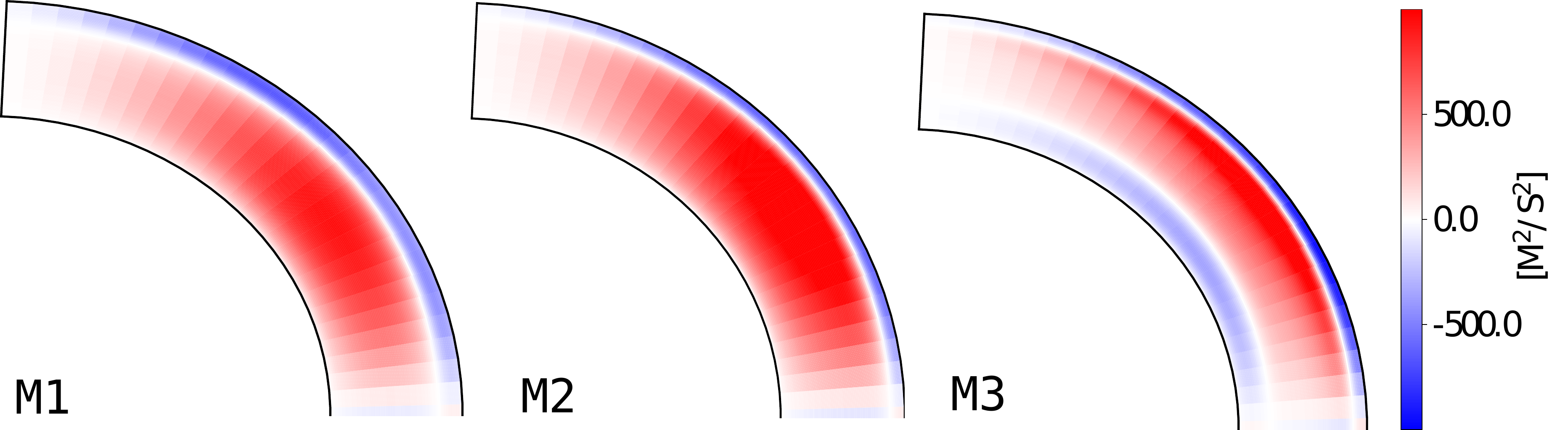}\caption{\label{fig:thphi}Distributions of $\left\langle {u'_{\theta}u'_{\phi}}\right\rangle $
in models M1, M2 and M3.}
}
\end{figure}

{Figure \ref{fig:thphi} shows the correlation of turbulent
velocities, $\left\langle {u'_{\theta}u'_{\phi}}\right\rangle $,
for the all three models. It is positive in the bulk of the convection
zone except the near-surface layer. Models M1 and M2 agree with earlier
conclusions of \citet{1993AA...279L...1K}. In model M3, this correlation
is negative near the bottom of the convection zone. The maximum of
$\left\langle {u'_{\theta}u'_{\phi}}\right\rangle $ is about $10^{3}~$m$^{2}$s$^{-2}$.
This is in agreement by an order of magnitude with the model of BY17.
In model M3, the correlation $\left\langle {u'_{\theta}u'_{\phi}}\right\rangle $
has a peak near the bottom boundary of the subsurface shear layer.
We compare this result with observations in the next section. }

{Our results suggest that the model of the meridional circulation
is sensitive to the choice of the radial Coriolis number profile in
the convection zone. In particular, we found that there is no sign
inversion of the $\Lambda$-effect for the profile of \citet{stix:02},
\[
\Omega^{*}=\Omega_{0}^{*}\left(\frac{x}{x_{0}}\right)^{7/6}\left(\frac{1-x}{1-x_{0}}\right)^{3/2},
\]
where $x=r/R_{\odot}$, and $\Omega_{0}^{*}$ is the Corioils number
at $x_{0}=0.8$ (e.g., $\Omega_{0}^{*}\equiv3$). Using this profile
and taking into account the inhomogeneity of the Coriolis number we
get no inversion of the $\Lambda$-effect. Results of that model are
very similar to our model M2. }

\section{Discussion and conclusions}

In this paper, we have presented a new model of the angular momentum
transport. {The model naturally explains the double-cell
meridional circulation structure. We confirm the conclusions
of \citet{2017ApJ835.9B} that the sign inversion of the radial part of
 the $\Lambda$-effect in the solar convection zone results in the
 double-cell meridional circulation.}

Our model demonstrated that the double-cell meridional
circulation pattern can result from the radial gradient of the Coriolis
number, which leads to the sign inversion of the nondissipative angular
momentum flux (the so-called $\Lambda$-effect) in the low part of
the solar convection zone. In our model, both the radial and
horizontal components of the $\Lambda$-effect change sign in the
lower part of the convection zone. The vector field of the nondiffusive
angular momentum transport in our most complete model M3 converges
in the middle of the convection zone. Therefore, the resulting profile
of the $\Lambda$-effect is consistent with the gyroscopic pumping
arguments. {Despite the fact  that in this paper the viscous part of the gyroscopic pumping equation
is substantially different from the simplified version  of
\citet{2017ApJ835.9B} the resulting distribution of the $\Lambda$-effect
in the solar convection zone  is qualitatively similar.}

We conclude that the knowledge of the convective turnover time distribution
is very important for robustness of the mean-field model prediction
of the large-scale flow structure inside the solar convection zone.
Moreover, the mixing-length theory's estimation of the turbulent parameters,
like the characteristic time-scale, $\tau_{c}$; the mixing length,
$\ell$; and the convective velocity RMS ${u'}$ may not be sufficiently
accurate at the convective zone boundaries because of overshooting
and anisotropy of convection.{ }

{\citet{2016AdSpR58.1490P} studied the robustness of the meridional
circulation structure with respect to variations of the latitudinal
profiles of the radial }$\Lambda$-effect{. They found that
the small increase of the radial }$\Lambda$-effect{ toward the
equator can result to the multi-cell meridional circulation structure
as well.} {The obtained pattern of the meridional circulation
is different to results of \citet{2014ApJ789L7Z}. We can conclude
that the variations of the latitudinal profiles of the radial }$\Lambda$-effect{
are less important to reproduce the results of helioseismology. To
understand the nature of the multi-cell meridional circulation in the stellar
convection zone the further theoretical progress in the theory of the
$\Lambda$-effect is needed.} 

The original theoretical formulation was obtained by \citet{KR93L}
using the second-order correlation approximation (SOCA,
\citealt{KR80})\textit{
  for the forced isothermal turbulence rather than for the stellar turbulent
convection}. The numerical simulations of \citet{kap2011} suggest
that in general case the SOCA can be valid only by an order of magnitude.
Another need for the future development follows from direct numerical
simulations. They often show multiple meridional circulation cells,
see, e.g., \citet{2016ApJ819.104G} or {\citet{2016arXiv160103730W}.
}These results are difficult to explain within our model. In our paper,
we discussed components of the $\Lambda$-effect, which contribute
to the generation of the azimuthal large-scale flow. In addition to them,
there is a theoretical possibility for the $\Lambda$-effect components
which generate the meridional circulation. This effect results from
the nondissipative part of the off-diagonal turbulent stresses: $\hat{T}_{r\theta}=\left\langle {u'_{\theta}u'_{r}}\right\rangle -{\displaystyle \frac{1}{4\pi\overline{\rho}}}\left\langle {b'_{\theta}b'_{r}}\right\rangle $,
where $\boldsymbol{{u}}'$ and $\boldsymbol{{b}}'$ are the fluctuating
velocity and magnetic field. The origin of this effect can be tightly
related to inhomogeneities of the kinetic and magnetic helicities
of turbulent flows \citep{2016PhRvE93c3125Y}. It is interesting that the
so-called ``anisotropic kinetic alpha - effect'' or the ``AKA-effect''
\citep{1987PhyD28.382F,1996GApFD..83..119P,2001AA379.1153B} also
results in the nondissipative part of the $\hat{T}_{r\theta}$ (as
well as the azimuthal components of $\mathbf{\hat{T}}$). {Therefore,
the theory of the $\Lambda$-effect, which is employed in our paper
is rather incomplete.}

{The model correlation $\left\langle {u'_{\theta}u'_{\phi}}\right\rangle $
can be compared with motions of sunspots \citep{2017SoPh292.86S}
and the large-scale magnetic fields \citep{1993SoPh146.401L}. Both of
these measurements show the positive correlation in the northern hemisphere
of the Sun. The sign of $\left\langle {u'_{\theta}u'_{\phi}}\right\rangle $
is the main reason for the solar-like differential rotation with the
equator rotating faster than the poles \citep{1989drsc.book.....R}.
Model M3 has the positive $\left\langle {u'_{\theta}u'_{\phi}}\right\rangle $
in the middle of the convection zone of the northern hemisphere of
the Sun. By comparing the model results with the measurements of $\left\langle {u'_{\theta}u'_{\phi}}\right\rangle $
from sunspot motions and the large-scale magnetic field tracers, we
can conclude that these features are likely rooted just below the
subsurface shear layer. Model M3 excludes positive values of $\left\langle {u'_{\theta}u'_{\phi}}\right\rangle $
for the magnetic field anchored at the bottom of the convection zone.
These conclusions agree with results of \citet{Beal1999} on rotation
of newly emerged active regions. }

In summary, the double-cell meridional circulation on the Sun is naturally
explained in our model because of a concurrent effect of the density
stratification and variations of the Coriolis force acting on the cyclonic
convection. {The key point is that the variation of the convective
turnover time with depth result in inversion of the sign of the non-dissipative
turbulent stresses (the $\Lambda$-effect) in the lower part of the
convection zone. However, the properties of the turbulent angular
momentum transport employed in this paper using the mean-field hydrodynamics
approach require further studies with the help of observations and
numerical simulations.}

Acknowledgments. Valery V. Pipin thanks the grant of Visiting Scholar
Program supported by the Research Coordination Committee, National
Astronomical Observatory of Japan (NAOJ). Also, a support
from of RFBR grants 16-52-50077 and 17-52-53203, and support of project
II.16.3.1 of ISTP SB RAS are greatly acknowledged. Alexander Kosovichev
thanks a support of NASA's grants NNX 14AB70G and NNX 17AE76A

\section{Appendix}

\subsection{The turbulent stress tensor}

Expression of the turbulent stress tensor results from the mean-field
hydrodynamics theory (see, \citealt{1994AN....315..157K,kit2004AR})
as follows: 
\begin{equation}
{\hat{T}_{ij}=\overline{\rho}\left\langle u_{i}u_{j}\right\rangle ,}\label{eq:stres-1}
\end{equation}
where ${u}$ is fluctuating velocity. Application the mean-field hydrodynamic
framework leads to the Taylor expansion given by Eq(\ref{eq:tstr}).
The viscous part of the azimuthal components of the stress tensor
is determined following \citealt{1994AN....315..157K} in the following
form: 
\begin{eqnarray}
{T_{r\phi}^{(\nu)}} & {=} & -{\overline{\rho}\nu_{T}\left\{ \Phi_{\perp}+\left(\Phi_{\|}-\Phi_{\perp}\right)\mu^{2}\right\} r\frac{\partial\sin\theta\Omega}{\partial r}}\label{eq:trf}\\
 & - & {\overline{\rho}\nu_{T}\sin\theta\left(\Phi_{\|}-\Phi_{\perp}\right)\left(1-\mu^{2}\right)\frac{\partial\Omega}{\partial\mu}}\nonumber \\
{T_{\theta\phi}^{(\nu)}} & {=} & {\overline{\rho}\nu_{T}\sin^{2}\theta\left\{ \Phi_{\perp}+\left(\Phi_{\|}-\Phi_{\perp}\right)\sin^{2}\theta\right\} \frac{\partial\Omega}{\partial\mu}}\label{eq:ttf}\\
 & + & {\overline{\rho}\nu_{T}\left(\Phi_{\|}-\Phi_{\perp}\right)\mu\sin^{2}\theta r\frac{\partial\Omega}{\partial r}},\nonumber 
\end{eqnarray}
where the eddy viscosity, $\nu_{T}$, is determined from the mixing-length
theory assuming the turbulent Prandl number ${Pr}_{T}={\displaystyle \frac{3}{4}}$:
\[
\nu_{T}=\frac{3\ell^{2}}{8}\sqrt{-\frac{g}{2c_{p}}\frac{\partial\overline{{s}}}{\partial r}}.
\]
The viscosity quenching functions, $\Phi_{\perp}$ and $\Phi_{\parallel}$,
depend nonlinearly on the Coriolis number, $\Omega^{*}=2\Omega\tau_{c}$
and they are determined by \citet{1994AN....315..157K}.

The nondiffusive flux of angular momentum can be expressed as follows
\citep{1989drsc.book.....R}: 
\begin{eqnarray}
\Lambda_{r\phi\phi}\overline{U}_{\phi} & = & r\Lambda_{V}\Omega\sin\theta,\nonumber \\
\Lambda_{V} & = & \nu_{T}\left(V^{(0)}+\sin^{2}\theta V^{(1)}\right),\label{eq:lv}\\
\Lambda_{\theta\phi\phi}\overline{U}_{\phi} & = & r\Lambda_{H}\Omega\cos\theta,\nonumber \\
\Lambda_{H} & = & \nu_{T}\sin^{2}\theta H^{(1)}\label{eq:lh}
\end{eqnarray}
The basic contributions to the $\Lambda$-effect are due to the density
stratification and the Coriolis force. \citet{KR93L} found the following
expression for the $\Lambda$-tensor coefficients: 
\begin{eqnarray}
V^{(0,\rho)} & = & \frac{r}{\rho^{2}}\frac{\partial}{\partial r}\left(\frac{1}{r}\frac{\partial}{\partial r}\tau^{2}\left\langle \boldsymbol{u}'^{2}\right\rangle \rho^{2}\left(I_{1}-I_{2}\right)\right)\label{eq:v0-1}\\
 & + & \tau^{2}\left\langle \boldsymbol{u}'^{2}\right\rangle \rho^{2}\left(I_{3}-I_{4}\right)r\frac{\partial}{\partial r}\frac{1}{r\rho}\frac{\partial\rho}{\partial r}\nonumber \\
 & + & \frac{1}{\rho^{3}}\frac{\partial\rho}{\partial r}\frac{\partial}{\partial r}\tau^{2}\left\langle \boldsymbol{u}'^{2}\right\rangle \rho^{2}\left(I_{5}-I_{6}\right),\nonumber 
\end{eqnarray}
\begin{eqnarray}
V^{(1,\rho)} & = & \frac{r}{\rho^{2}}\frac{\partial}{\partial r}\left(\frac{1}{r}\frac{\partial}{\partial r}\tau^{2}\left\langle \boldsymbol{u}'^{2}\right\rangle \rho^{2}I_{2}\right)\label{eq:v1-1}\\
 & + & \tau^{2}\left\langle \boldsymbol{u}'^{2}\right\rangle \rho^{2}I_{4}r\frac{\partial}{\partial r}\frac{1}{r\rho}\frac{\partial\rho}{\partial r}\nonumber \\
 & + & \frac{1}{\rho^{3}}\frac{\partial\rho}{\partial r}\frac{\partial}{\partial r}\tau^{2}\left\langle \boldsymbol{u}'^{2}\right\rangle \rho^{2}\left(I_{5}-I_{6}\right),\nonumber 
\end{eqnarray}
where $I_{n}$ are functions of the Coriolis number and $H^{(1,\rho)}=-V^{(1,\rho)}$.
If we neglect the contributions of all inhomogeneities except the
density stratification and variations of the Coriolis number in functions
$I_{n}$ we get the following results: 
\begin{eqnarray}
V^{(0,\rho)} & = & \left(\frac{\ell}{H_{\rho}}\right)^{2}\left(J_{0}+J_{1}\right)-\frac{\ell^{2}}{H_{\rho}}\frac{\partial}{\partial r}\left\{ \left(J_{0}+J_{1}\right)-I_{5}+I_{6}\right\} +\ell^{2}\frac{\partial^{2}}{\partial r^{2}}\left(I_{1}-I_{2}\right),\label{eq:v0ro}\\
V^{(1,\rho)} & = & -\left(\frac{\ell}{H_{\rho}}\right)^{2}J_{1}+\frac{\ell^{2}}{H_{\rho}}\frac{\partial}{\partial r}\left\{ J_{1}+I_{6}\right\} +\ell^{2}\frac{\partial^{2}}{\partial r^{2}}I_{2},\label{eq:v1ro}
\end{eqnarray}
where, $H_{\rho}=-{\displaystyle \left(\frac{\partial\log\bar{\rho}}{\partial r}\right)^{-1}}$
, $J_{0}=4I_{1}+2I_{5}$ and $J_{1}=-4I_{2}-2I_{6}$ are defined in
the above cited paper. Functions $I_{n}$ have the following form
(also, see, \citealt{KR93L}): 
\begin{eqnarray*}
I_{1} & = & \frac{1}{4\Omega^{*4}}\left(\frac{6+5\Omega^{*2}}{1+\Omega^{*2}}-\left(6+\Omega^{*2}\right)\frac{\arctan\Omega^{*}}{\Omega^{*}}\right),\\
I_{2} & = & \frac{1}{8\Omega^{*4}}\Bigl(60+\Omega^{*2}-\frac{6\Omega^{*2}}{1+\Omega^{*2}}\\
 & + & \left(\Omega^{*4}-15\Omega^{*2}-60\right)\frac{\arctan\Omega^{*}}{\Omega^{*}}\Bigr),\\
I_{3} & = & \frac{1}{2\Omega^{*4}}\left(-3+\frac{\Omega^{*2}}{1+\Omega^{*2}}+3\frac{\arctan\Omega^{*}}{\Omega^{*}}\right),\\
I_{4} & = & \frac{1}{2\Omega^{*4}}\left(-15+\frac{2\Omega^{*2}}{1+\Omega^{*2}}+\left(15+3\Omega^{*2}\right)\frac{\arctan\Omega^{*}}{\Omega^{*}}\right),\\
I_{5} & = & \frac{1}{2\Omega^{*4}}\left(-3+\left(\Omega^{*2}+3\right)\frac{\arctan\Omega^{*}}{\Omega^{*}}\right),\,I_{6}=\frac{1}{2}I_{4}.
\end{eqnarray*}

Using the mixing-length approximation for the adiabatic distribution
of the thermodynamic parameters we define the mixing-length parameter
as follows: ${\displaystyle \frac{\ell}{H_{\rho}}\equiv\frac{\alpha_{MLT}}{\gamma}}$.
As was noted by \citealt{KR93L} in case of $\Omega^{*}\ll1$ we have:
${\displaystyle I_{1}=\frac{1}{30}}$, ${\displaystyle I_{3}=-\frac{1}{5}}$,
${\displaystyle I_{5}=\frac{1}{15}}$, and $I_{2,4,6}=0$. Therefore
the derivatives of $I_{n}$ and $J_{0,1}$ in Eq(\ref{eq:v0ro}) and
Eq(\ref{eq:v1ro}) can be neglected.

For case of fast rotation, it is found $I_{2}={\displaystyle \frac{\pi}{16\Omega^{*}},}$
$J_{1}={\displaystyle -\frac{\pi}{4\Omega^{*}},}$ and others functions
are of $O\left(\Omega^{*-3}\right)$. Then for $V^{(0,\rho)}$ from
Eq(\ref{eq:v0ro}) we have 
\begin{eqnarray}
V^{(0,\rho)} & \approx & \frac{\ell}{H_{\rho}}J_{1}\left(\frac{\ell}{H_{\rho}}+\ell\frac{\partial\log\Omega^{*}}{\partial r}\right),\label{eq:v0r2}\\
V^{(1,\rho)} & \approx & -\frac{\ell}{H_{\rho}}J_{1}\left(\frac{\ell}{H_{\rho}}+\ell\frac{\partial\log\Omega^{*}}{\partial r}\right)
\end{eqnarray}
where the first term is positive , and the second term is negative
(as the $\Omega^{*}$ decreased toward surface) . The first term dominate
in the upper part of the convection zone but the second term increases
steeply toward the bottom, and overcome the first one (Fig. \ref{fig:lamb}a)
and the density height scale varies slowly. We conclude that it is
important to keep the second contributions in Eq(\ref{eq:v0ro}) and
Eq(\ref{eq:v1ro}).

In the final expressions of the $\Lambda$-effect we take into account
anisotropy of convective velocities (see, \citealt{kit2004AR}). Therefore,
the final coefficients of the $\Lambda$-tensor are: 
\begin{eqnarray}
V^{(0)} & = & \Bigl[\left(\frac{\alpha_{MLT}}{\gamma}\right)^{2}\left\{ J_{0}+\!J_{1}+\!a\left(I_{0}+\!I_{1}\right)\right\} \label{v0-f}\\
 & - & \Bigl(\frac{\alpha_{MLT}\ell}{\gamma}\frac{\partial}{\partial r}\left\{ \left(J_{0}+J_{1}\right)-I_{5}+I_{6}\right\} +\ell^{2}\frac{\partial^{2}}{\partial r^{2}}\left(I_{1}-I_{2}\right)\Bigr],\nonumber \\
V^{(1)} & = & -\left\{ \left(\frac{\alpha_{MLT}}{\gamma}\right)^{2}\left(J_{1}+aI_{1}\right)-\frac{\alpha_{MLT}\ell}{\gamma}\frac{\partial}{\partial r}\left(J_{1}+I_{6}\right)-\ell^{2}\frac{\partial^{2}}{\partial r^{2}}I_{2}\right\} ,\label{v1-f}
\end{eqnarray}
and ${H^{(1)}=-V^{(1)}}$. We employ the parameter of the turbulence
anisotropy $a={\displaystyle \frac{\overline{{u_{h}^{2}}}-2\overline{{u_{r}^{2}}}}{\overline{{u_{r}^{2}}}}=}2$,
where ${u_{h}}$ and ${u_{r}}$ are the horizontal and vertical RMS
velocities \citep{kit2004AR}.

The first RHS term of Eq.(\ref{eq:vort}) describes dissipation of
the mean vorticity, $\omega$. Similarly to \citet{rem2005ApJ} we
approximate it as follows, 
\begin{equation}
-{\left[\boldsymbol{\nabla}\times\frac{1}{\overline{\rho}}\boldsymbol{\nabla\cdot}\overline{\rho}\hat{\mathbf{T}}\right]_{\phi}\approx2\nu_{T}\phi_{1}\left(\Omega^{*}\right)\nabla^{2}\omega,}\label{eq:vort2-1}
\end{equation}
where the rotational quenching function $\phi_{1}$ is given by \citet{1994AN....315..157K}. We have tried
to apply a more general formalism including all components of the
eddy-viscosity tensor for rotating turbulence provided by \citet{1994AN....315..157K},
and obtained results that are similar to the approximation given by
Eq(\ref{eq:vort2-1}).
\end{document}